\documentclass{article}

\usepackage[english]{babel}
\usepackage{pdfpages}

\usepackage[letterpaper,top=2cm,bottom=2cm,left=3cm,right=3cm,marginparwidth=1.75cm]{geometry}

\usepackage{amsmath}
\usepackage{amsthm}
\usepackage{amsfonts}
\usepackage{graphicx}
\usepackage{bm}
\usepackage{xcolor}
\usepackage{setspace}
\usepackage[colorlinks=true, allcolors=blue]{hyperref}
\usepackage[inline]{enumitem}
\usepackage{rotating}
\usepackage{multirow}
\usepackage{longtable}
\usepackage{placeins}
\usepackage{authblk}

\usepackage[normalem]{ulem} 

\usepackage[backend=biber,url=false,sorting=none,style=numeric-comp]{biblatex}
\DeclareMathOperator*{\argmin}{arg\,min}

\def\br{{\bm r}}
\def\bR{{\mathbf{R}}}

\def\bA{{\bm A}} 
\def\bB{{\mathcal{B}}}
\def\bv{{\bm v}}
\def\bc{{\bm c}}
\def\bb{{\rm b}}
\def\ee{{\rm e}}
\def\Nat{{N_{\mathrm{at}}}}
\def\Ne{{N}}
\def\Nd{{N_{\mathrm{data}}}}
\def\P{\mathcal{P}}
\def\fcut{{f_{\rm cut}}}
\def\fcutb{{f_{\rm cut}^\bb}}
\def\fcute{{f_{\rm cut}^\ee}}
\def\Oiii{{{\rm O}(3)}}

\newtheorem{theorem}{Theorem}
\newtheorem{proposition}[theorem]{Proposition}

\usepackage{xr} 

\makeatletter
\newcommand*{\addFileDependency}[1]{
  \typeout{(#1)}
  \@addtofilelist{#1}
  \IfFileExists{#1}{}{\typeout{No file #1.}}
}
\makeatother

\newcommand*{\myexternaldocument}[1]{
    \externaldocument{#1}
    \addFileDependency{#1.tex}
    \addFileDependency{#1.aux}
}

\myexternaldocument{SI}

\title{A symmetry-preserving and transferable representation for learning the Kohn-Sham density matrix}
\author[1]{Liwei Zhang}
\author[2]{Patrizia Mazzeo}
\author[3]{Michele Nottoli}
\author[2]{Edoardo Cignoni}
\author[2]{Lorenzo Cupellini} 
\author[3]{Benjamin Stamm}
\affil[1]{Institut f\"ur Geometrie und Praktische Mathematik, RWTH Aachen University, Templergraben 55, 52062 Aachen, Germany}
\affil[2]{Dipartimento di Chimica e Chimica Industriale, Università di Pisa, 56124 Pisa, Italy}
\affil[3]{Universität Stuttgart, Institute of Applied Analysis and Numerical Simulation, Pfaffenwaldring 57, 70569 Stuttgart, Germany}
\date{}

\begin{document}
\maketitle

\begin{abstract}
The Kohn-Sham (KS) density matrix is one of the most essential properties in KS density functional theory (DFT), from which many other
physical properties of interest can be derived. In this work, we present a parameterized representation for learning the mapping from a molecular configuration to its corresponding density matrix using the Atomic Cluster Expansion (ACE) framework, which preserves the physical symmetries of the mapping, including isometric equivariance and Grassmannianity. Trained on several typical molecules, the proposed representation is shown to be systematically improvable with the increase of the model parameters and is transferable to molecules that are not part of and even more complex than those in the training set. 
The models generated by the proposed approach are illustrated as being able to generate reasonable predictions of the density matrix to either accelerate the DFT calculations or to provide approximations to some properties of the molecules. 

\end{abstract}

\section{Introduction}

Computational chemistry often deals with many quantum mechanical calculations repeated on the same system or on similar systems. Examples are molecular dynamics (MD) simulations, repeated calculations on a statistical sampling, geometry optimizations, or even scans along some interesting coordinate. In all these cases, the results of already performed calculations can be used to fit a machine learning model able to predict energies and properties of subsequent calculations\cite{Fedik2022}.

In the context of quantum chemistry, machine learning models have been used to fit properties, for example, the energy\cite{Qiao2020,Shao2023,Chen2020,Dick2020,Christensen2021,Welborn2018} and atomic forces\cite{Unke2021MLFF,Pinheiro2021,Unke2021Spooky}, or to predict more fundamental quantities like the Hamiltonian\cite{Schtt2019,Zhang2022,Nigam2022,Li2022,Cignoni2024,Shakiba2024} and the wavefunction\cite{Hermann2020,Li2022_fermionic}.
Machine learning models have also been used directly as interatomic potentials for molecular dynamics simulations of a variety of systems\cite{Chmiela2019,Zinovjev2024,Galvelis2023,Kabylda2024,Mazzeo2024}.

Among the more fundamental quantities, various methods have been proposed to fit the electronic density matrix. These either target the electronic density in real space\cite{Brockherde2017,Alred2018,Chandrasekaran2019,Gong2019,Fabrizio2019,CuevasZuvira2020,Meyer2020,Ellis2021,CuevasZuvira2021,Jrgensen2022,Focassio2023,Lee2024}, or they target the corresponding electronic density matrix in a basis\cite{Grisafi2018,Lewis2021,Briling2021,Rackers2023,Shao2023,Hazra2024}.
Fitting the electronic density is a powerful strategy, as the density can then be directly used to compute different observables. Other strategies often need to train an ad hoc model for each property of interest, but multiple properties are often required for answering a scientific question.

While being less general than fitting in real space, fitting the density in a suitable basis removes any projection error and removes the barrier between the predicted density and the quantum chemistry package of choice, which can be used to compute the properties of interest. 
Fitting the electronic density matrix provides two additional advantages. First, in the context of Hartree-Fock or density functional theory (DFT), an electronic density matrix can simply be used as an initial guess for the upcoming self-consistent field (SCF) calculation, instead of directly using it to access properties. This hybrid approach represents a middle ground between a full SCF calculation and directly using the density to access the properties: it retains the full accuracy of a normal SCF procedure, but at a reduced computational cost\cite{Hazra2024}. The better the guess, the more efficient is the full accuracy model.
Second, given a predicted electronic density matrix $D$, it is possible to assemble the corresponding Fock / Kohn-Sham matrix $F(D)$, and the commutator $FD - DF$ provides a measure of how accurate the prediction is, thus providing the opportunity to either discard low quality predictions or mark the data points with the worst predictions, which is useful in active learning strategies\cite{VDO2023hyperactive}. 

However, the required mapping from the molecular configurations (coordinates and atomic numbers) to the corresponding density matrices is in general complicated and of high dimensionality, and therefore difficult to learn. The fitting problem becomes treatable by introducing appropriate molecular descriptors, which take into account physical knowledge such as invariance or equivariance of the target property. In this way, the required design space can be reduced. 
More specifically, the descriptors are functions of the molecular parameters satisfying a series of requisites: they are desired to be injective (exactly or approximately) and economical to compute, and should capture the aforementioned symmetries of the target property.
Depending on the order in which the various invariances are introduced, different classes of descriptors are obtained. A possible strategy is to compute translationally and rotationally invariant functions of the coordinates, and only then introducing the permutational invariance. Examples of descriptors of this kind are permutationally invariant polynomials (PIPs)\cite{Xie2009} and their variant atomic permutationally invariant polynomials (aPIPs)\cite{vdOord2020}.
Alternatively, it is possible to compute functions that are permutationally and translationally invariant, thereafter enforcing the rotational invariance. This is the strategy followed by the smooth overlap of atomic positions (SOAP)\cite{Bartk2013}, by the atomic cluster expansions (ACE)\cite{Drautz2019} and by the Behler-Parrinello descriptors\cite{Behler2011,Behler2021}.
The ACE descriptors are of particular interest as they include, in principle, many-body terms of arbitrarily high order, and are cheaper to compute than other alternatives\cite{Lysogorskiy2021}. Notably, it has also been generalized to capture the equivariant properties\cite{Drautz2020,Zhang2022}.

In this contribution, we propose a strategy that combines the strengths of the equivariant ACE descriptors with the flexibility of fitting the electronic density matrix in a basis, which respects the intrinsic properties of the density matrix. Specifically, the electronic density matrix is approximated with a linear regression in an ACE basis, similarly to the previous work on self-consistent Hamiltonians from one of the authors\cite{Zhang2022}. 
The strategy is used to train both specific models (that is, trained on a single molecule) and unified models (trained on multiple molecules).
The resulting models are systematically improvable and, in the case of unified models, also transferrable to unseen molecules, provided that they share some chemical similarity with the training set.
Both the specific and unified models can be used to reduce the number of SCF iterations or to directly predict the properties of interest.

\section{Methods}

\subsection{Density Matrix}
Let $\bR = \{(Z_I, \br_I)\}_{I=1}^\Nat:= \{\sigma_I\}_{I=1}^\Nat$ be a molecular configuration consisting of $\Nat$ atoms and $\Ne$ (valence) electron-pairs, with $Z_I\in\mathbb{N}$ and $\br_I\in\mathbb{R}^3$ characterizing the atomic number and the position of the $I$-th atom, respectively. The union of all $Z_I$ characterizes the different elements in this given system, whose cardinality will be denoted by $n$. Other properties of the atoms can potentially also be included in $\sigma_I$ but that would go beyond the scope of this paper. If an $N_g(\ge\Ne)$ dimensional discretization space is adopted, in which the orbitals are approximated, then the corresponding KS equation will read as

\begin{equation*} \label{eq:kse0}
    F_{\bR}[D_{\bR}]C_{\bR} = S_{\bR}C_{\bR}E_{\bR}.
\end{equation*}
where $F_{\bR}$ and $S_{\bR}\in\mathbb{R}^{N_g\times N_g}$ are the discretized KS operator (Hamiltonian) and the overlap matrix respectively, $C_{\bR}\in\mathbb{R}^{N_g\times N}$ represents the coefficients of the orbitals in a given basis, $D_{\bR} = C_{\bR}C_{\bR}^T$ is the density matrix, the main object of this paper, and $E_{\bR}$, a diagonal matrix of order $\Ne$, which contains the $\Ne$ corresponding eigenvalues of the system (sorted in ascending order). Without loss of generality, we assume that $S_{\bR} = I_{N_g}$, by adopting the L\"owdin orthonormalization\cite{Lowdin1956} if necessary. Under this setting, the above equation \eqref{eq:kse0} can be rewritten as 
\begin{equation} \label{eq:kse}
    F_{\bR}[D_{\bR}]C_{\bR} = C_{\bR}E_{\bR}.
\end{equation}
If $C_{\bR}$ is chosen to be orthonormal, then $D_{\bR}$ should lie in the following manifold

\begin{equation}\label{eq:Grassmann}
\mathcal{G}_{N_g}^\Ne :=\{D\in\mathbb{R}^{N_g\times N_g}: D^2=D^T=D, \text{tr}(D)=\Ne\},
\end{equation}
which is equivalent to an $(N_g, N)$-Grassmann manifold, hence our notation. 

\begin{figure}[h]
\centering
\includegraphics[width=1.0\textwidth]{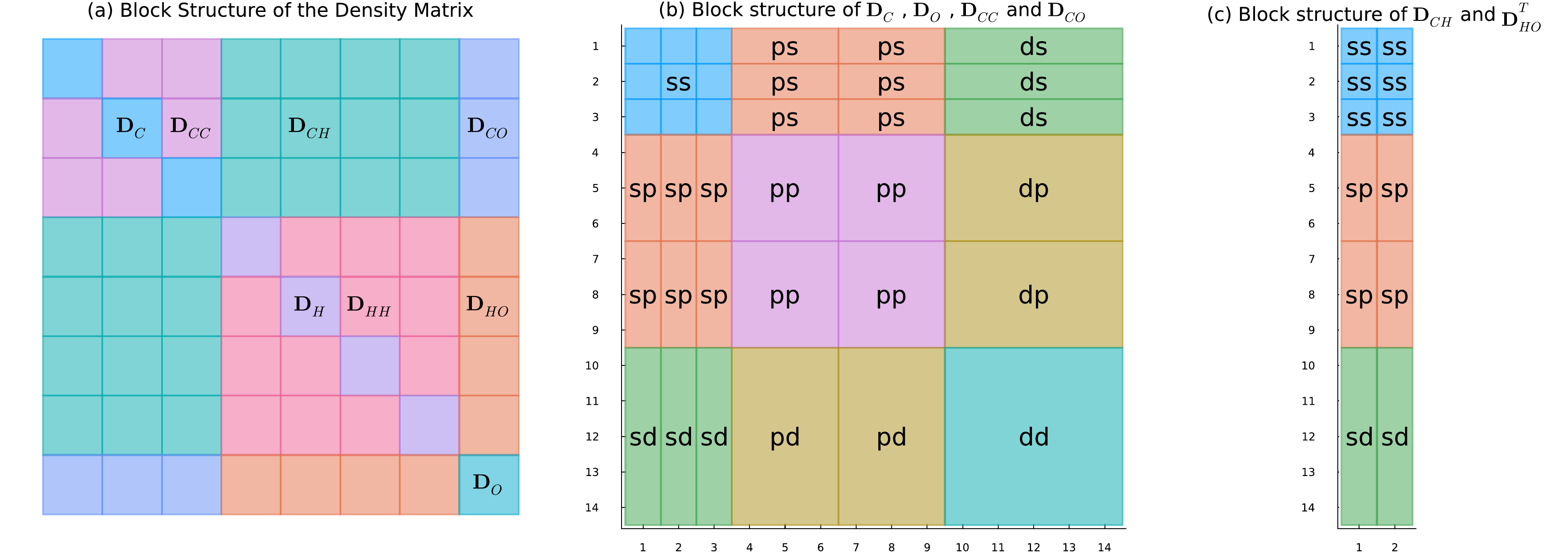}
\caption{Block structure of the density matrix of the $C_3H_4O$ molecule, where the atomic basis 6-31G(d) is used (the block structure of $D_{HH}$ is omitted as it is just a 2 by 2 matrix consisting of 4 ss-blocks).}
\label{Fig:DMStructure}
\end{figure}

In the context of linear combinations of atomic orbitals (LCAO), the discretization space is spanned by a set of atomic orbitals $\{\phi_{I\alpha}\}_{I\in\{1,\ldots,\Nat\},\alpha\in\mathcal{I}_{Z_I}}$ where $\mathcal{I}_{Z_I}$ is the index set of the atomic orbitals centered at the $I$-th atom, depending only on the atomic number $Z_I$. The density matrix, consequently, has elements that are invariant under translations of the system or permutations of the index of the atoms, 
and is equivariant under rotations and reflections of the whole system. It can be divided into several subblocks that have respective symmetries and can be learned independently. In Figure \ref{Fig:DMStructure}, we take $C_3H_4O$ as an example to illustrate the block structure of the density matrix. Note that a similar strategy is used in Ref.~\cite{Zhang2022} \and is here extended to a case with multiple different elements. The detailed derivation of the block-wise equivariance of the density matrix can be found in Appendix \ref{app:equiv}. For simplicity, we omit the subscript $\bR$ in $D_\bR$ hereafter when no ambiguity is introduced. 

As can be seen from Figure \ref{Fig:DMStructure}(a), there are two types of blocks appearing in the density matrix, the diagonal blocks and the off-diagonal ones. Depending on contexts, they are also called \textit{onsite} and \textit{offsite}, or \textit{homo-} and \textit{hetero-orbital}, respectively. We will use the terms \textit{onsite} and \textit{offsite} throughout this paper. 
For the targeted systems having $n$ different elements, there exist, 
$n(n+3)/2$ matrix-valued functions which correspond to interactions of distinct elements (there can be a lack thereof, for instance, when the system has only one oxygen atom, there is no O-O interaction). 
As such, the block of the density matrix corresponding to the $I$-th and $J$-th atom is of the form
\begin{equation}\label{eq:DIJ1}
    D_{IJ} = 
    \begin{cases}
        D_{Z_I}(\bR_I), & I=J, \\
        D_{(Z_I,Z_J)}(\bR_{IJ}),  & I\ne J, ~ Z_I\le Z_J, \\
        D_{(Z_J,Z_I)}(\bR_{JI})^T, & I\ne J, ~ Z_I > Z_J,
    \end{cases}
\end{equation}
where $\bR_I$ and $\bR_{IJ}$ are global configurations, translated in order to be centered at the $I$-th atom or at some specific point of the $(I,J)$-th bond (the line segment that connects the two atoms) respectively. Note that the \textit{Hermitian-ness} of the density matrix (\textit{i.e.}, $D = D^T$) is used in the last line of \eqref{eq:DIJ1}. To unify the notations, we sometimes use the convention that $\bR_{II} = \bR_I$. 

In addition, each of such matrix-valued functions can be further divided into completely independent sub-blocks corresponding to the atomic orbitals, as shown in Figure \ref{Fig:DMStructure}(b) and \ref{Fig:DMStructure}(c), \textit{i.e.},

\begin{equation}\label{eq:DZIZJab}
\begin{aligned}
     D_{Z_I}(\bR_I) &= [ D_{Z_I}^{\alpha\beta}(\bR_I)]_{\alpha, \beta\in\mathcal{I}_{Z_I}},\\
     D_{(Z_I,Z_J)}(\bR_{IJ}) &= [ D_{(Z_I,Z_J)}^{\alpha\beta}(\bR_{IJ})]_{\alpha\in\mathcal{I}_{Z_I},\beta\in\mathcal{I}_{Z_J}}.
\end{aligned}
\end{equation}
Our target then becomes the unified functionals $D_{{Z}_i/({Z}_i,{Z}_j)}^{\alpha\beta}$ for various atomic numbers $Z_i,~Z_j$, which have their distinct isometric equivariance and can be dealt with separately. Such structure of the density matrix forms the foundation of the transferability and parallelizability of the proposed method. We refer readers to Appendix \ref{app:equiv} for a detailed discussion. 

\begin{figure}[h]
\centering
\includegraphics[width=1.0\textwidth]{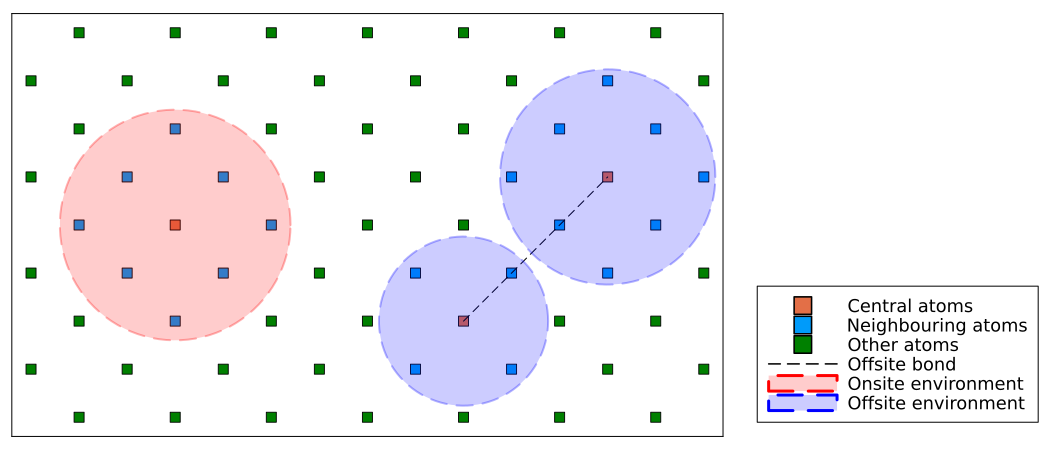}
\caption{Illustration of the local atomic environments. Atoms in the red sphere form an \textit{onsite} local configuration, with the radius of the sphere being the onsite cutoff. Atoms in the blue spheres form the \textit{offsite} local configuration, with the radii of the sphere being the offsite cutoffs that can be chosen differently from the \textit{onsite} one. }
\label{Fig:loc_env}
\end{figure}

In practice, it is commonly believed that only atoms near the central atoms make substantial contributions to the corresponding part of the density matrix (also known as the nearsightedness of the object). As a result, certain cutoff strategies are often used when constructing the input atomic environment. We illustrate our truncation strategy in Figure \ref{Fig:loc_env}, where the particles in the red and blue spheres form the \textit{onsite} and \textit{offsite} environments $\bR_I$ and $\bR_{IJ}$, respectively. The rigorous definitions of the truncated $\bR_I$ and $\bR_{IJ}$ are provided in Appendix \ref{app:equiv}. 

Here, we assume that atoms of the same element are discretized by the same set of bases. However, the method proposed in this work can potentially be extended to the more general setting where atoms of the same element are assigned different atomic orbital basis functions, simply by artificially treating them as having different atom types.




\subsection{Representation of the density matrix}
One of the goals of this paper is to provide a faithful representation of the density matrix, respecting its inherent physical symmetries as much as possible to facilitate its learning. To this end, we adopt the equivariant ACE descriptors\cite{Drautz2019, Zhang2022} to approximate the functions $D_{\bullet}$, where the symbol $\bullet$ can be one of the indices appearing in the right hand side of \eqref{eq:DZIZJab}. 

For each function $D_{\bullet}$, there exists a set of ACE bases $\{\bB_{\bullet,\bv}\}_\bv$ as functions of the local environments $\bR_{I}$ or $\bR_{IJ}$, which has the same isometric equivariance as $D_{\bullet}$ and asymptotically spans the function space to which $D_{\bullet}$ belongs\cite{Dusson2022}. The size of the basis is determined merely by two parameters: (i) the correlation order $\nu$, which corresponds to the body order in physics (up to a constant, precisely, it is $1$ and $2$ less for \textit{onsite} and \textit{offsite}, respectively) and (ii) the maximum polynomial degree $d_{\rm{max}}$, indicating the resolution of the one-particle basis. Additional details about these two parameters and the corresponding ACE basis, as well as the definition of the one-particle basis, can be found in Appendix \ref{app:ACEBasis}. Given the basis $\{\bB_{\bullet,\bv}\}_\bv$, we can approximate each function $D_{\bullet}$ by a linear combination of $\{\bB_{\bullet,\bv}\}_\bv$: 
\begin{equation}\label{eq:DIJab}
    D_{\bullet} \approx \sum_\bv c_{\bullet,\bv}\bB_{\bullet,\bv}.
\end{equation}

To predict the corresponding sub-block of the density matrix, the only thing left now is to estimate the coefficient $c_{\bullet,\bv}$ for all possible indices $\bullet$.

\subsection{Parameter estimation} 
Suppose that a dataset is given of the form $\{(\bR^{(k)},D^{(k)})\}_k$, where $k$ is the index of the data point, $\bR^{(k)}$ is the $k$-th (global) molecular configuration and $D^{(k)}$ is the corresponding density matrix. The dataset can first be transformed into sets of local atomic clusters and their corresponding portions of the density matrix, according to the atomic number of each atom in $\bR^{(k)}$, as
\[
    \{\big(\bR^{(k)}_{IJ}, D^{(k)}_{\bullet}\big)\}_{k,I,J},
\]

where the subscript $\bullet$ has the same meaning as that in the preceding subsection. These sets are then used to train the coefficients of the corresponding models \eqref{eq:DIJab} independently. One of the most direct ways to estimate the coefficients is through a least squares approach, that is, they are determined by minimizing 
\begin{equation}\label{eq:LS}    
    L(\bc_{\bullet}) = \sum_{k,I,J}\|D^{(k)}_{\bullet} - \sum_\bv c_{\bullet,\bv} \bB_{\bullet,\bv}(\bR^{(k)}_{IJ})\|^2 + \lambda\|\Gamma_{\bB_{\bullet}}\|^2,
\end{equation}
where $\bc_{\bullet} = \{c_{\bullet,\bv}\}_\bv$, 
$\Gamma_{\bB_{\bullet}}$ refers to some Tikhonov regularizer that can be customized with respect to the basis ${\bB_{\bullet}}$, and $\lambda$ is a regularization parameter. Throughout our experiments, we use $\lambda=10^{-4}$ and choose $\Gamma$ to be the smooth prior given in Ref.~\cite{Zhang2022}. Once an (approximate) minimizer is found, it is possible to provide an approximation of the ground state density matrix through \eqref{eq:DIJab} for any given configuration $\bR$ as long as its chemical composition of elements does not go beyond that of the training set.

\subsection{Retraction}

The construction of the ACE basis as well as our representation \eqref{eq:DIJab} ensure that the predicted density matrix $D$ has the desired isometric-equivariance. However, it does not guarantee that the prediction belongs to the Grassmann manifold \eqref{eq:Grassmann}, \textit{i.e.} that it is a valid density matrix. To bring this restriction back, we introduce a retraction operator that maps $D$ to the manifold.

Since $D$ is a real symmetric matrix of size $N_g\times N_g$, its eigenvalue decomposition can be written as
\[
    D = U_D \Sigma_D U_D^T,
\]
where $U_D\in\mathbb{R}^{N_g\times N_g}$ is unitary and $\Sigma_D$ is a diagonal matrix containing all the eigenvalues of $D$, sorted in descending order. The retraction is then defined as 
\begin{equation}\label{eq:eigen_retract}
    \P(D) = U_D E_{N_g}^N U_D^T,
\end{equation}
where 
\[
    E_{N_g}^N = 
    \begin{bmatrix}
        I_N & {\bm 0} \\
        {\bm 0} & {\bm 0}
    \end{bmatrix}
    \in\mathbb{R}^{N_g\times N_g}.
\]
We mention that after applying the retraction, $\mathcal{P}(D)$ remains isometric equivariant and is the nearest element on the Grassmann manifold to $D$. A proof of this statement, as well as the well-definedness of $\P$, is given in Appendix \ref{app:proof1}.

\begin{figure}[h]
\centering
\includegraphics[width=1.0\textwidth]{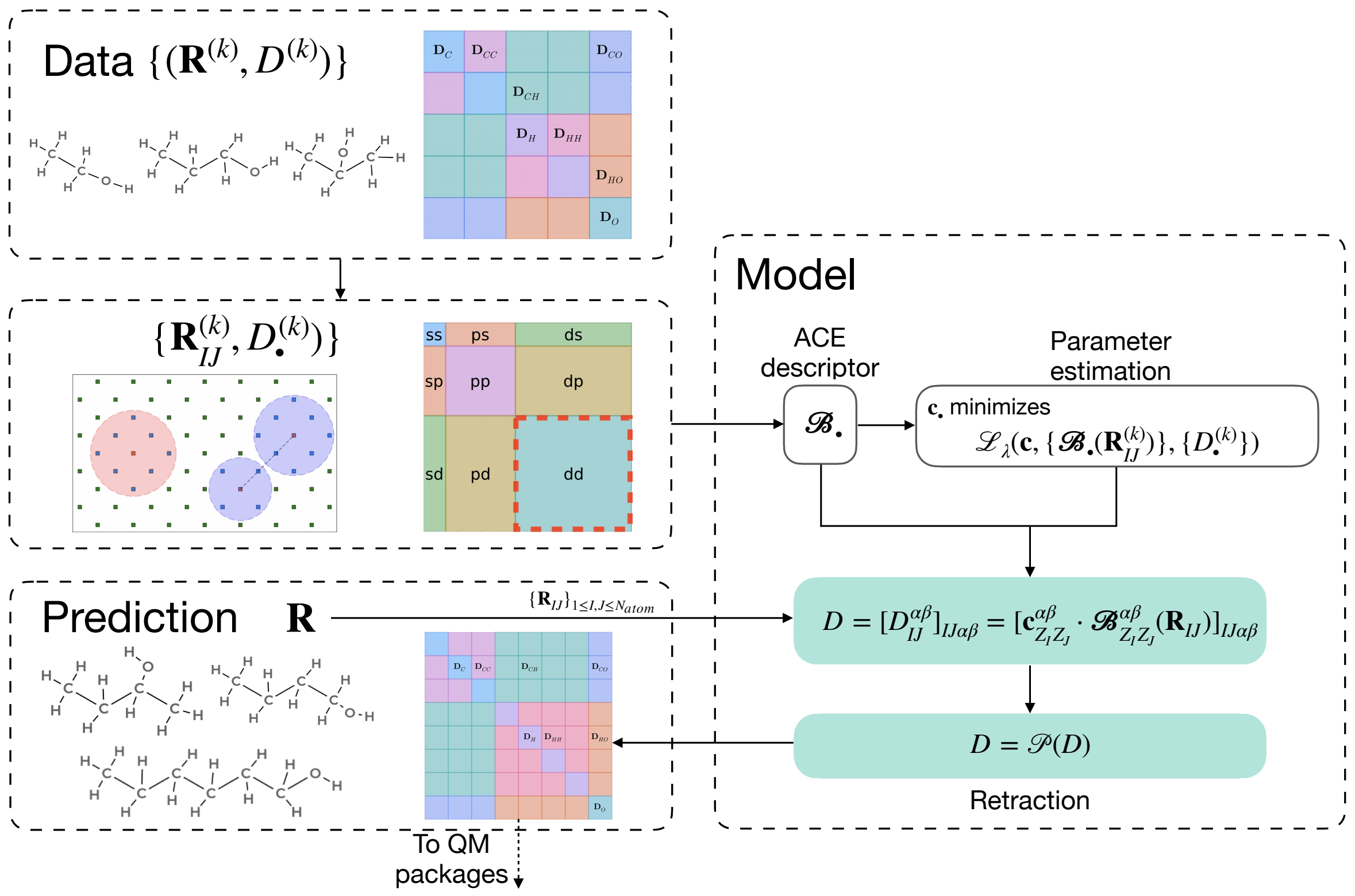}
\caption{A schematic of the process of learning the density matrix described in this paper. Here, the loss function $\mathcal{L}_\lambda$ is defined as \eqref{eq:LS}. The molecules for which the density matrices are predicted can be different from those in the training set. Finally, the predicted density matrix can be directly sent to some quantum chemistry packages of choice for further operations.}
\label{Fig:DM-schematic}
\end{figure}

The following schematic (Figure \ref{Fig:DM-schematic}) summarizes the whole procedure of our density matrix prediction scheme, from which we can see that the whole procedure, apart from the retraction step, is essentially parallelizable, including training and prediction. 

\FloatBarrier 

It is worth noting that the proposed scheme does not require specific data origination, but just requires a consistent atomic basis discretization across the data set (or more abstractly, it requires only the equivariance of the data).
The resulting density matrix can be used in many different application scenarios (cf. Section \ref{sec:applications}).  We remark that a similar strategy is used in \cite{Zhang2022} to fit the self-consistent Hamiltonian (\textit{i.e.} Kohn-Sham) matrix for periodic crystal systems. We compared the aforementioned approach of learning the Hamiltonian matrix therein with the method proposed in this work, whose results can be found in Appendix \ref{app:DMvsHM}.

\section{Results and discussion}\label{sec:result}

Our approach of learning the density matrix was tested on various systems described with DFT. We designed tests of increasing complexity to validate the approach and test its performance. The various tests required both training and test datasets generated using a standard protocol, whose details can be found in Section~\ref{subsec:data}. In general, the performance of the various models was assessed by computing the Root-Mean-Square-Error(RMSE) between the references ($\{D_\text{ref}^{(k)}\}_{k=1}^{\Nd}$) and predicted density matrices ($\{D_\text{pred}^{(k)}\}_{k=1}^{\Nd}$) as
\begin{equation}\label{eq:rmse}
{\text{RMSE}}_D =  \sqrt{\frac{\sum_{k=1}^\Nd \|D_{\text{ref}}^{(k)} - D_{\text{pred}}^{(k)}\|_F^2}{\sum_{i=1}^\Nd \big(N_g^{(k)}\big)^2}},
\end{equation}  
where $N_g^{(k)}$ is the size of the $k$-th density matrix. 

\subsection{Data preparation}\label{subsec:data}

Each dataset was prepared with the same protocol, consisting of a sampling step and a QM calculation step.
In the sampling step, each molecule was optimized with DFT B3LYP/6-31G in water, treated with IEFPCM\cite{Cances1997}, and solvated with an octahedral box of TIP3P waters\cite{Mark2001}, extending up to 35~\AA~from the molecule. The solvent was then minimized while keeping the molecule fixed. Thereafter, the whole system was heated from 0~K to 100~K in a 5~ps NVT simulation and from 100~K to 310~K in a 100~ps NPT simulation. The QM/MM production simulation was then run for 150~ps in the NVT ensemble, using the Langevin thermostat. The molecule was treated at the DFTB3 level of theory\cite{Gaus2011} with 30b-3-1 parameters\cite{Gaus2014,Lu2015}. 
Electrostatic interactions were treated with PME\cite{Darden1993}, using a 10 $\AA$ cutoff to divide the direct and reciprocal space. The first 50~ps of production trajectory were discarded. All simulations were performed with AMBER\cite{Case2022}.

In the second step, QM DFT calculations were run on equally spaced frames along the trajectory, using the  $\omega$B97XD/6-31G(d) level of theory, and enforcing the use of spherical atomic basis functions. 
The training-and-testing dataset comprises nine organic molecules featuring different functional groups, whereas the test-only datasets comprise nine similar but distinct molecules.
For training-and-testing datasets, the calculations were run on 10000 frames, whereas for test only datasets, the calculations were run only on 100 frames. All the calculations for the datasets were performed using Gaussian 16\cite{g16}.

The datasets were generated by storing the coordinates, the overlap matrices, the coefficient matrices, the Kohn-Sham matrices, as well as metadata such as the list of atoms and the calculation level in HDF5 binary files. An overview of the datasets is reported in Table \ref{tab:datasets}. All the datasets are available in the corresponding archive for the sake of reproducibility.

\begin{table}[]
    \centering
    \begin{tabular}{l|r} \hline
    \textbf{Molecule} & \textbf{N frames} \\ \hline
    Acetaldehyde & 10000 \\
    Acrolein & 10000 \\ \hline
    Aniline & 10000 \\
    o-Toluidine & 10000 \\
    m-Toluidine & 10000 \\
    Benzene & 100 \\
    Toluene & 100 \\
    Phenol & 100 \\
    Benzaldehyde & 100 \\ 
    p-Toluidine & 100 \\ \hline
    1-Propanol & 10000 \\
    1-Butanol & 10000 \\
    2-Butanol & 10000 \\
    1-Hexanol & 10000 \\
    Ethanol & 100 \\
    2-Propanol & 100 \\
    2-Hexanol & 100 \\
    1-Heptanol & 100 \\ \hline
    \end{tabular}
\caption{Datasets used in this work. Different molecules are grouped according to their chemical class: aldehydes, aromatics, alcohols. The level of theory is DFT $\omega$B97XD/6-31G(d).
}
\label{tab:datasets}
\end{table}

\subsection{Specific models}

To assess the method we proposed, we first show that it generates systematically improvable results, so that we can refrain from fine-tuning the choice of model parameters (\textit{i.e.}, correlation orders $\nu$ and maximum polynomial degrees $d_{\max}$). To this end, we show the results of molecule-specific models, each of which is trained with the geometries of only one molecule. For each training molecule (those with 10000 available frames in Table \ref{tab:datasets}) apart from acrolein and butanol, we use the first 3000 frames or a subset thereof for training, and test the resulting models on the rest among the 10000 frames. For the two exceptional molecules, we sampled alternatively/tertiarily from the first 6000/9000 frames, respectively. In any case, less than 30\% of available data points are included in the training. More details on the selection of the dataset can be found in Table S1 and Figure S5 in the Supporting Information (SI). The training dataset is then used to train the models with $\nu = 2,~3$ and $4\le d_{\max}\le 8$. For the local truncation, we choose the cutoffs (the radii of the three spheres displayed in Figure \ref{Fig:loc_env}, which are chosen uniformly in this work regardless of the atomic types) to be $4.0\mathring{A}$ and $6.5\mathring{A}$, respectively. For the sake of simplicity, we only show the results of aniline and propanol. In the following Figure \ref{Fig:D-MSE-AniProp}, the x-axis stands for the degree $d_{\max}$ used to generate the descriptors, whereas the y-axis is the element-wise RMSE of the predicted density matrix $D_{\text{pred}}$ defined in \eqref{eq:rmse}, displayed in the logarithmic scale. 
\begin{figure}[h]
\centering
\includegraphics[width=0.48\textwidth]{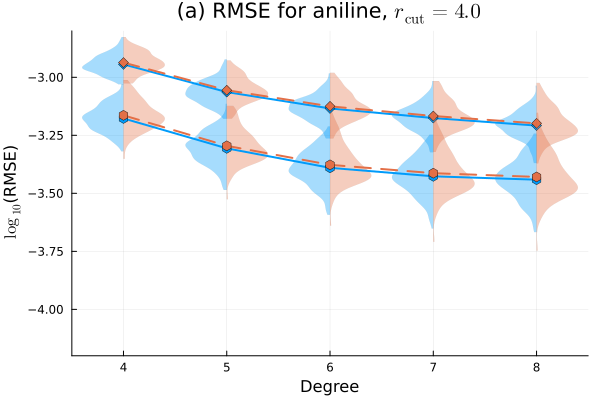}~~
\includegraphics[width=0.48\textwidth]{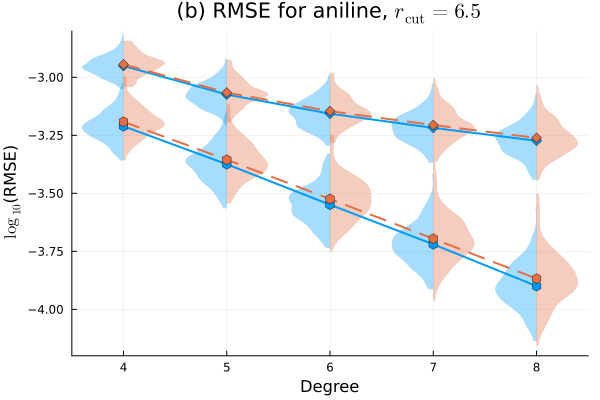}~~
\\
~~\includegraphics[width=0.48\textwidth]{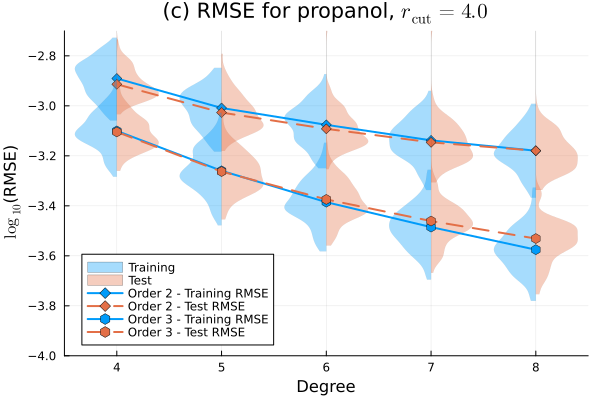}~~
\includegraphics[width=0.48\textwidth]{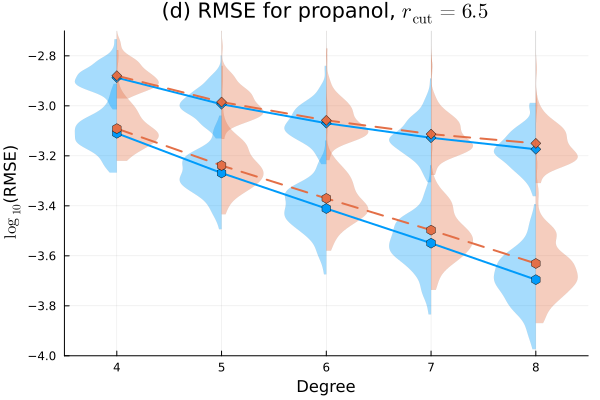}~~
\caption{The RMSEs~\eqref{eq:rmse} of the predicted density matrices for: (a) aniline with $r_{\text{cut}}$ = 4.0, (b) aniline with $r_{\text{cut}}$ = 6.5, (c) propanol with $r_{\text{cut}}$ = 4.0 and (d) propanol with $r_{\text{cut}}$ = 6.5, obtained by the corresponding specific models with respect to different degrees $d_{\max}$ and for various orders $\nu$. The solid and dashed lines refer to the average training and test set errors, and the shaded areas show the distribution of the errors for the corresponding models.}
\label{Fig:D-MSE-AniProp}
\end{figure}

As illustrated in Figure \ref{Fig:D-MSE-AniProp}, the training and test set RMSEs align with each other nicely, and are nearly normally distributed, even with less than $30\%$ of the data being involved in the training phase. The similarity of the training/test-set errors shows little overfitting in the training, validating the effects of the regularization we used (cf. equation \eqref{eq:LS}). Comparing the two truncation parameters, we find that the models trained with larger cutoff radii can give better results, but the differences are not too significant. It reaches the smallest average RMSE at around $10^{-4}$ and $2\cdot 10^{-4}$ for Aniline and Propanol, respectively, which corresponds to a relative error in the whole density matrix of around 0.2$\%$ to 0.35$\%$ (note that $\|D\|_F^2=\Ne$ the number of electrons, so the relative error is simply given by $\|D-D_{\text{pred}}\|_F/\sqrt{N}$). Furthermore, the RMSE, in all cases, monotonically decreases with increasing ACE basis size, which implies that smaller errors can be expected simply by increasing either of the two model parameters. 

\subsection{Unified model}\label{subsec:unified}

We then extend the training set to configurations from several distinct molecules. More specifically, the training set used in this subsection consists of a total of 2700 frames (300 frames evenly sampled from the first 9000 frames of each of the 9 training molecules). This extended training set is used to train the largest model mentioned above ($\nu = 3$, $d_{\max} = 8$) in order to obtain a unified model, which is then tested on the last 1000 configurations of the training molecules. 
We compare the average test-set RMSE obtained by the unified model with those obtained by the corresponding specific models of the same size, whose results can be found in the first two columns of Table \ref{tab:unified}. We note that to make the model more capable of capturing the similarity only of local chemical structures across different molecules, we choose $r_{\text{cut}}=4.0$ for the unified model. The results for the unified model with $r_{\text{cut}}=6.5$ are given in Table S1 in the SI, which indeed shows that a smaller cutoff is favorable in terms of the generalizability of the models. 

\begin{table}[]
\setstretch{1.4}
\centering
\begin{tabular}[h!]{c|ccc}
\hline
\textbf{Molecule} & \textbf{Specific Model} & \textbf{Unified Model} & \textbf{Unified Model-A} \\ \hline
Acetaldehyde        & 4.416$\cdot$10$^{-5}$       & 3.278$\cdot$10$^{-4}$     & 3.299$\cdot$10$^{-4}$ \\
Acrolein            & 2.514$\cdot$10$^{-4}$       & 5.028$\cdot$10$^{-4}$     & 5.081$\cdot$10$^{-4}$ \\
\hline
Aniline             & 4.300$\cdot$10$^{-4}$       & 4.868$\cdot$10$^{-4}$     & 4.876$\cdot$10$^{-4}$ \\
o-Toluidine         & 5.430$\cdot$10$^{-4}$       & 5.962$\cdot$10$^{-4}$     & 5.940$\cdot$10$^{-4}$ \\
m-Toluidine         & 5.384$\cdot$10$^{-4}$       & 5.824$\cdot$10$^{-4}$     & 5.822$\cdot$10$^{-4}$ \\
Benzene$^*$         & -                          & 4.058$\cdot$10$^{-4}$     & 3.646$\cdot$10$^{-4}$ \\
Toluene$^*$         & -                          & 6.369$\cdot$10$^{-4}$     & 5.980$\cdot$10$^{-4}$ \\
Phenol$^{**}$       & -                          & 4.809$\cdot$10$^{-3}$     & 6.770$\cdot$10$^{-4}$ \\
Benzaldehyde$^{**}$ & -                          & 4.129$\cdot$10$^{-3}$     & 1.201$\cdot$10$^{-3}$ \\
p-Toluidine$^{**}$  & -                          & 2.840$\cdot$10$^{-3}$     & 6.293$\cdot$10$^{-4}$ \\
\hline
1-Propanol          & 3.049$\cdot$10$^{-4}$       & 4.427$\cdot$10$^{-4}$     & 4.444$\cdot$10$^{-4}$ \\
1-Butanol           & 4.510$\cdot$10$^{-4}$       & 4.921$\cdot$10$^{-4}$     & 4.934$\cdot$10$^{-4}$ \\
2-Butanol           & 9.173$\cdot$10$^{-4}$       & 1.494$\cdot$10$^{-3}$     & 1.509$\cdot$10$^{-3}$ \\
1-Hexanol           & 1.031$\cdot$10$^{-3}$       & 5.324$\cdot$10$^{-4}$     & 5.314$\cdot$10$^{-4}$ \\
Ethanol$^*$         & -                          & 9.644$\cdot$10$^{-4}$     & 8.999$\cdot$10$^{-4}$ \\
2-Propanol$^*$      & -                          & 7.701$\cdot$10$^{-4}$     & 7.480$\cdot$10$^{-4}$ \\
2-Hexanol$^*$       & -                          & 7.384$\cdot$10$^{-4}$     & 7.353$\cdot$10$^{-4}$ \\
1-Heptanol$^*$      & -                          & 8.896$\cdot$10$^{-4}$     & 8.980$\cdot$10$^{-4}$ \\
\hline
\end{tabular}
\caption{The test set RMSEs obtained by the (3,8)-models trained on different datasets ($r_{\text{cut}}$ = 4.0). There is no specific model for test-only molecules, hence some dashes in the first column. In particular, the test molecules with a superscript $*$ are not included in the training process at all, and those with $**$ are involved in the training of the augmented model, Unified Model-A, with only 10 frames each included in training.} 
\label{tab:unified}
\end{table}

As indicated in the first two columns of Table \ref{tab:unified}, the unified model overall achieves a performance comparable to the specific models trained on each molecule independently. This indicates that the model is able to gather information from distinct molecules, and offers the advantage of predicting the density matrices for multiple systems within a single model. Whereas the RMSE of the unified model for acetaldehyde is slightly higher, the unified model performs even better than the specific one for 1-hexanol. This demonstrates that the models generated by the proposed method are able to predict the density matrices of diverse molecular systems, despite their inherent structural dissimilarities, as long as a certain number of their frames are included in the training set. We remark that the training set of the unified model comprises a slightly smaller number of configurations than that of the specific models, and was sampled without particular strategies.

For reference, we also trained a unified model for the alcohols, which is a simpler task compared to the unified one we introduce in this subsection. The results for the Unified Alcohols Model can be found in table S1 in the SI. 

\subsection{Transfer to other molecules}

As a more challenging task, we directly use the unified model obtained in subsection 3.3 to predict the density matrices of some molecules beyond the training set, which may be larger or more complex. Specifically, we test the models on those molecules having 100 frames in Table \ref{tab:datasets}. 
The corresponding average test set RMSEs are reported also in the second column of Table \ref{tab:unified}, from which we observe that the unified model can provide faithful predictions of the density matrices for benzene, toluene and all the alcohols, for which the obtained errors are similar to those within the training process. However, the unified model struggles with giving good predictions for phenol, benzaldehyde, and is less well in predicting the density matrices for p-toluidine. The poor prediction on these molecules can be attributed to the lack of information in the training set. Indeed, while the dataset includes both carbonyl compounds and alcohols, the chemical behaviour of these functional groups changes significantly when they are bonded to aromatic molecules. Additionally, when an aromatic molecule has two substituents, their effect depends on their relative positions. This explains why, despite the inclusion of o- and m-toluidine in the training set, the model struggles to accurately predict for p-toluidine.

To test whether the weaker performance of the unified model on the three molecules is caused by the limitation of the method itself or just by the training set, we design an augmented training set, which consists of the data points of the previous unified training set, and 10 frames from each of the three molecules, evenly sampled from the first 90 frames (2730 configurations in total). We train the above (3,8)-model with the augmented training set, and obtain a new model Unified Model-A, with the suffix ``A" indicating that it is trained with an augmented set. The augmented unified model is again used to predict the density matrices for all the involved molecules, and the corresponding test set RMSEs are listed in the third column of Table \ref{tab:unified}. The results show that the augmented unified model achieves a higher accuracy for the three molecules with previously critical accuracy, which is similar in magnitude to that of the training molecules, while maintaining a comparable effectiveness for the other systems involved. 

The RMSE results presented in this section suggest that the models generated by the proposed method can be uniformly refined simply by increasing the two model parameters. In addition, the proposed unified models can be transferred to the molecules that are not known at the training stage, provided some similarity in the chemical geometries. The performance of the generated models is mainly limited by the design of the training set, rather than the representation itself. 

\section{Applications}\label{sec:applications}

In this section, we showcase how the predicted density matrices can be used in some specific scenarios, as extended quality tests for the proposed models. 
\subsection{Accelerating the SCF iterations}

A natural application to try is to use the predictions as the initial guesses of the SCF procedure, as it is no matter what not of full accuracy. For each test geometry, we use the proposed models to predict the density matrix and provide it to Gaussian as an initial guess. For these calculations, we used the development version of the Gaussian suite of programs\cite{gdv}. Communication with Gaussian is possible thanks to the GauOpen open-source library\cite{gauopen}.
We compared the number of iterations required to achieve convergence with all our models and with the default guess available in Gaussian. Table~\ref{table:alliter} reports the average of iterations obtained with the default guess, specific models with $r_\text{cut} = 4.0$ and unified models for SCF convergence tolerance $10^{-6}$. The same results for the convergence levels $10^{-7}$ and $10^{-8}$ are reported in Table~S4 in the SI. The value within parentheses indicates the percentage of reduction with respect to the default guess. 
The table shows that, as expected, the specific model achieves the highest reduction in the number of iterations for each molecule, with a maximum for acetaldehyde, where a 44\% reduction is observed. We also compared the performance of the specific models with $r_\text{cut} = 4.0$ and $r_\text{cut} = 6.5$, finding no significant differences (see Table S2 in the SI). 
On average, the specific models allow us to save three iterations (30\%).
Moving to the unified model, we observe that for the majority of the molecules in the training set, the predicted density is comparable to that obtained with the respective specific model, with a slightly greater loss of accuracy for the two carbonyl molecules. 
For what concerns the out-of-sample molecules, the model exhibits good transferability for alcohols, achieving comparable results for both known and unknown molecules. Conversely, the predictions for phenol, p-toluidine, and benzaldehyde were particularly poor, even falling below the accuracy of the default guess. However, as demonstrated by the RMSEs, including just 10 frames in the training set for these three molecules enhances the performance and reduces the number of iterations by around two (Unified Model-A).

It is worth mentioning that the computational time for predicting a density matrix for a given configuration using our model is almost negligible compared to a single SCF iteration. For example, it takes about 112 ms to obtain a predicted density matrix for a propanol molecule using the unified model in a single thread, whereas a single SCF iteration, even carried out on 6 threads, takes an average of 626 ms. Hence, the percentage of reduction is almost exactly the acceleration that we gain.  

\begin{table}[h!]
\setstretch{1.3}
\begin{tabular}{c|cccc}
\hline
   \textbf{Molecule} &      \textbf{Default guess}     &   \textbf{Specific Model}     &  \textbf{Unified Model}  & \textbf{Unified Model-A}\\ \hline
Acetaldehyde &  9.4 $\pm$ 0.1       & 5.2 $\pm$ 0.1 ($\sim$  44\%)          & 7.5 $\pm$ 0.1 ($\sim$  20\%)          & 7.5 $\pm$ 0.1 ($\sim$  20\%)          \\ 
Acrolein &  10.5 $\pm$ 0.1          & 7.5 $\pm$ 0.2 ($\sim$  29\%)          & 8.3 $\pm$ 0.2 ($\sim$  21\%)          & 8.3 $\pm$ 0.2 ($\sim$  21\%)          \\ 
\hline
Aniline &  9.9 $\pm$ 0.1            & 7.2 $\pm$ 0.1 ($\sim$  28\%)          & 7.5 $\pm$ 0.1 ($\sim$  24\%)          & 7.5 $\pm$ 0.1 ($\sim$  24\%)          \\ 
o-Toluidine &  10.0 $\pm$ 0.0       & 7.6 $\pm$ 0.1 ($\sim$  24\%)          & 7.8 $\pm$ 0.1 ($\sim$  22\%)          & 7.8 $\pm$ 0.1 ($\sim$  22\%)          \\ 
m-Toluidine &  10.0 $\pm$ 0.0       & 7.3 $\pm$ 0.1 ($\sim$  27\%)          & 7.6 $\pm$ 0.1 ($\sim$  24\%)          & 7.6 $\pm$ 0.1 ($\sim$  24\%)          \\ 
Benzene$^*$ &  9.0 $\pm$ 0.0            &   -    & 7.4 $\pm$ 0.1 ($\sim$  18\%)          & 7.4 $\pm$ 0.1 ($\sim$  18\%)          \\ 
Toluene$^*$ &  9.0 $\pm$ 0.0            &     -       & 8.0 $\pm$ 0.0 ($\sim$  11\%)          & 7.9 $\pm$ 0.1 ($\sim$  12\%)          \\ 
Phenol$^{**}$ &  9.9 $\pm$ 0.1             &   -         & 10.0 $\pm$ 0.1 ($\sim$  -1\%)         & 8.1 $\pm$ 0.1 ($\sim$  18\%)          \\ 
Benzaldehyde$^{**}$ &  10.7 $\pm$ 0.1      &   -    & 10.5 $\pm$ 0.1 ($\sim$   2\%)         & 9.0 $\pm$ 0.0 ($\sim$  16\%)          \\ 
p-Toluidine$^{**}$ &  10.0 $\pm$ 0.0       &   -    & 8.3 $\pm$ 0.1 ($\sim$  16\%)          & 7.8 $\pm$ 0.1 ($\sim$  21\%)          \\ 
\hline
1-Propanol &  9.0 $\pm$ 0.0         & 6.0 $\pm$ 0.1 ($\sim$  33\%)          & 6.5 $\pm$ 0.1 ($\sim$  27\%)          & 6.5 $\pm$ 0.1 ($\sim$  28\%)          \\ 
1-Butanol &  9.0 $\pm$ 0.0          & 6.4 $\pm$ 0.1 ($\sim$  29\%)          & 6.6 $\pm$ 0.1 ($\sim$  27\%)          & 6.5 $\pm$ 0.1 ($\sim$  27\%)          \\ 
2-Butanol &  9.0 $\pm$ 0.0          & 7.2 $\pm$ 0.1 ($\sim$  20\%)          & 7.0 $\pm$ 0.1 ($\sim$  22\%)          & 7.1 $\pm$ 0.1 ($\sim$  22\%)          \\ 
1-Hexanol &  9.0 $\pm$ 0.0          & 6.4 $\pm$ 0.1 ($\sim$  29\%)          & 6.5 $\pm$ 0.1 ($\sim$  28\%)          & 6.5 $\pm$ 0.1 ($\sim$  28\%)          \\ 
Ethanol$^*$ &  9.1 $\pm$ 0.0            &   -    & 7.2 $\pm$ 0.1 ($\sim$  21\%)          & 7.1 $\pm$ 0.1 ($\sim$  21\%)          \\ 
2-Propanol$^*$ &  9.0 $\pm$ 0.0         &   -    & 7.1 $\pm$ 0.1 ($\sim$  21\%)          & 7.0 $\pm$ 0.0 ($\sim$  22\%)          \\ 
2-Hexanol$^*$ &  9.0 $\pm$ 0.0          &   -    & 7.0 $\pm$ 0.1 ($\sim$  23\%)          & 7.0 $\pm$ 0.1 ($\sim$  22\%)          \\ 
1-Heptanol$^*$ &  9.0 $\pm$ 0.0         &   -    & 6.6 $\pm$ 0.1 ($\sim$  27\%)          & 6.6 $\pm$ 0.1 ($\sim$  27\%)          \\ 
\hline
\end{tabular}
\caption{Average number of SCF iterations obtained by the (3,8)-models trained with different datasets ($r_\text{cut}$ = 4.0). The values reported within parentheses indicate the percentage of reduction with respect to the default Gaussian guess. The test molecules with a superscript $*$ are not included in the training process at all, and those with $**$ are involved in the training of Unified Model-A, with only 10 frames each included. }
\label{table:alliter}
\end{table}

\subsection{Predictions of physical properties}
\begin{figure}[h!]
    \centering
    \includegraphics[width=0.9\linewidth]{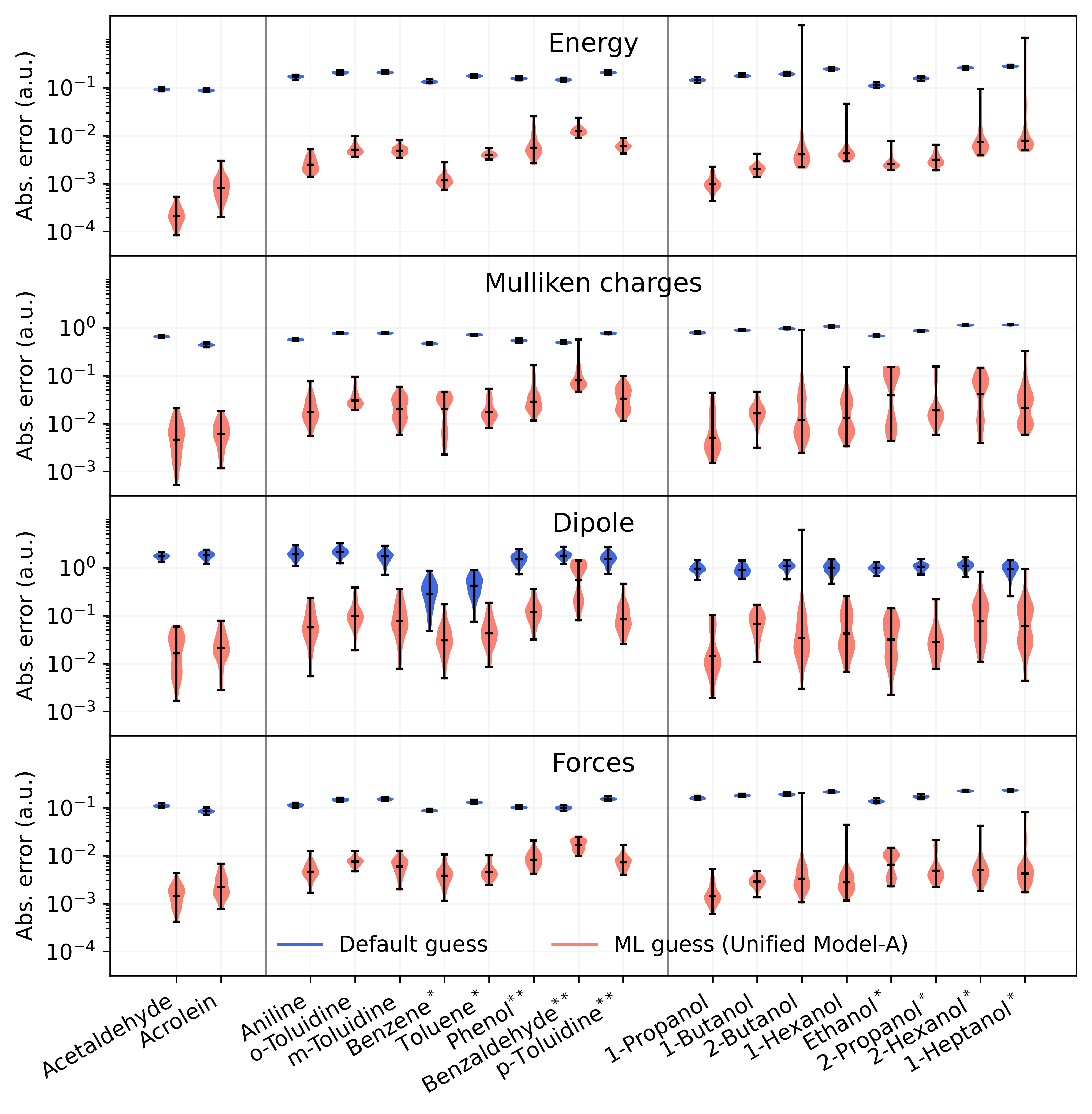}
    \caption{Plot of the error (in logarithmic scale) for energy, Mulliken charges, dipole moment and forces after a single SCF cycle. The blue line represents the default guess provided by Gaussian, while the pink line corresponds to the density matrix predicted using Unified Model-A. The test molecules with a superscript $*$ are not included in the training process at all, and those with $**$ are involved in the training of Unified Model-A, with only 10 frames each included. }
    \label{fig:obs_unified_A}
\end{figure}

The predicted density matrices can also be directly used to derive physical properties of interest, obtaining satisfactory predictions. 
This was achieved by providing the predicted density matrix as a guess and forcing Gaussian to stop the SCF procedure after a single iteration. Figure~\ref{fig:obs_unified_A} illustrates the error in energy, Mulliken charges, dipole moment, and forces with respect to the results obtained from the corresponding converged density matrix. 
The plot compares the errors obtained using the density matrix predicted with Unified Model-A (pink) and the default guess available in Gaussian (blue), for which we also ran a single SCF iteration for consistency. 
The same plots for specific models, unified alcohols model and unified model are reported in the SI (see Figures S1, S2 and S3).

Averaging over all molecules, we obtain a mean absolute error (MAE) of 2.7 kcal/mol for energy and 6.5 kcal $\cdot$ mol$^{-1}$ $\cdot$ \AA$^{-1}$ for forces. Although they do not achieve chemical accuracy (1 kcal/mol for energies and 1 kcal $\cdot$ mol$^{-1}$ $\cdot$ \AA$^{-1}$ for forces) except for the two aldehydes, the predictions can be considered as qualitatively correct results in most of the cases. Typically, the average errors of the properties derived from the predicted density matrix for all the involved molecules are 1 to 3 orders of magnitude smaller than those from the Gaussian default guesses. This trend holds consistently for both the aldehyde and aromatic families.  Despite the existence of some outliers for the alcohols, especially 2-butanol, which also turned out to be the one having the largest test set RMSE within the training molecules (unified models), they are only rare occurrences, as indicated by the error distribution shown in the violin plots. We expect that this can be resolved by adjusting the training set to include the structures corresponding to the outliers. This  result also suggests that one may need to give the alcohol family more weight in the training. It is therefore likely that a better selection of training points, obtained for example by active learning approaches (see \textit{e.g.} Ref.~\cite{VDO2023hyperactive} and the references therein), will give more stable errors. In Section~\ref{sec:commutators}, we provide a potential way to determine whether a given prediction should be disregarded or whether the corresponding structure should be included in the training set to improve model performance.

\subsection{Commutator and errors}\label{sec:commutators}
\begin{figure}[h!]
    \centering
    \includegraphics[width=0.5\linewidth]{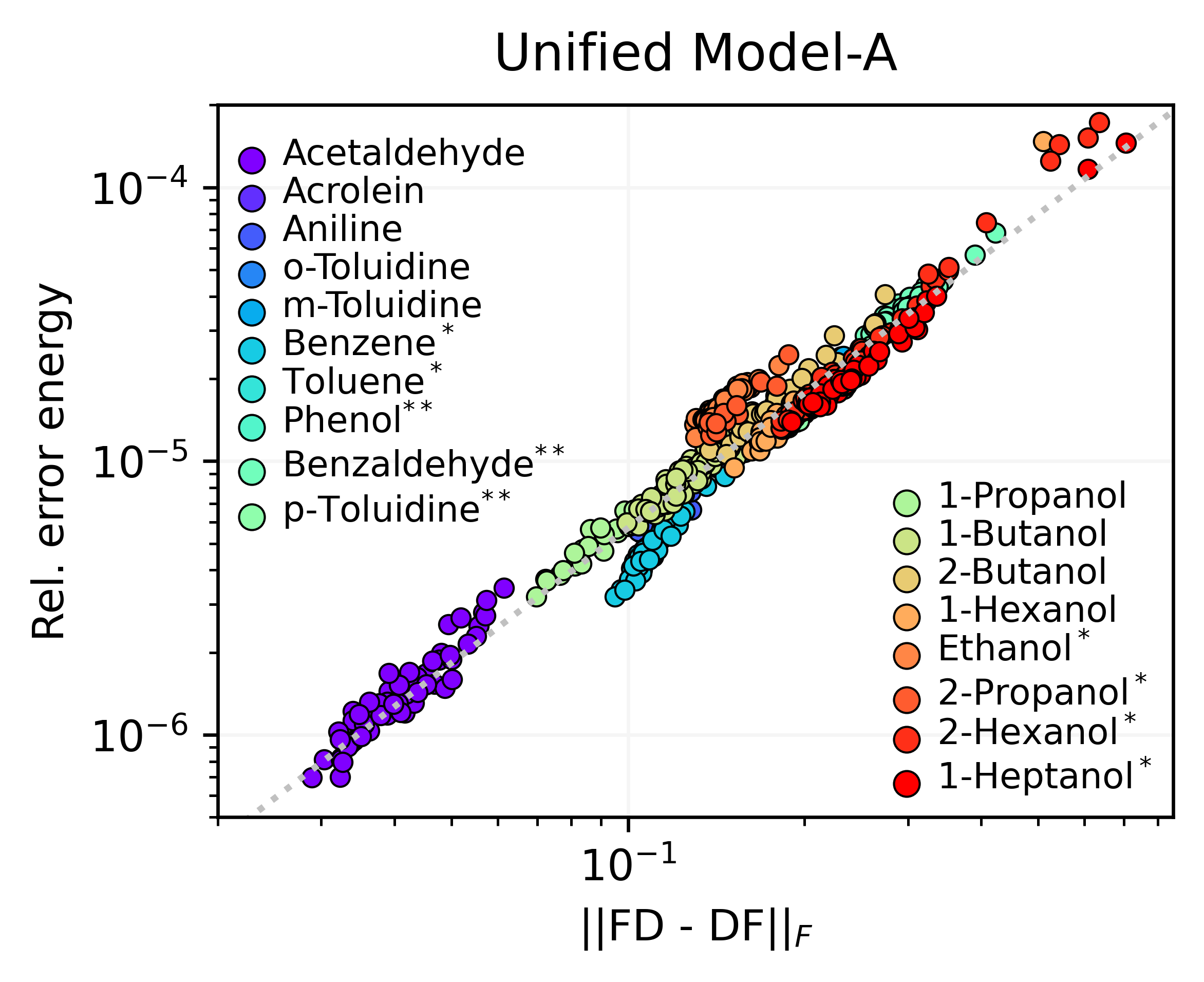}
    \caption{Plot of the Frobenius norm of the commutator between $F$ and $D$ versus the relative error in energy obtained after a single SCF cycle, using the density matrix predicted by the Unified Model-A as a guess. The test molecules with a superscript $*$ are not included in the training process at all, and those with $**$ are involved in the training of Unified Model-A, with only 10 frames each included.}
    \label{fig:comm:ene_unified_A}
\end{figure}

As a last application, we use the predicted density matrices to compute the corresponding KS matrix $F=F(D)$, and check how well the commutator condition $FD = DF$ is fulfilled. 
Indeed, when convergence is achieved, $FD = DF$ must hold exactly. Therefore, the norm of $FD - DF$ is a residue and can serve as a physical parameter to evaluate the accuracy of the prediction. In Figure~\ref{fig:comm:ene_unified_A}, we present the relationship between the commutator violation error, measured in the Frobenius norm, and the relative error in the predicted energy. It turns out that there is an empirical algebraic relation observed between the two errors. Similar plots for other properties are provided in the SI (Figure~S4), which also demonstrate positive correlations while the trend is less clear compared to that for the energies. Thus, we can use the commutator error to determine whether to disregard a prediction, without accessing the real physical properties of interest. From another perspective, we can also use the commutator error as an indicator of which frames to include in the training process in an active learning framework. As shown in the previous section, it is indeed the design of the training set that limits the accuracy of the proposed method, and this is therefore one of our immediate future works. 

\FloatBarrier

\section{Conclusions}

We presented a simple yet powerful regression model for learning the ground-state density matrix of arbitrary molecules in an atom-centered basis set. Our model exploits the flexibility and the favorable symmetry properties of the equivariant ACE descriptors, which represent a natural set of features to represent 
the density matrix.
The resulting model can be improved systematically by increasing the size of the ACE basis (order and degree) or by tuning the training samples. More importantly, our model can learn the relationship between molecular geometry and density matrix using information from multiple distinct molecules. This opens the possibility of building unified models for molecules with similar local structure, in contrast to other approaches\cite{Shao2023,Hazra2024}.
As a consequence, our model is transferable to unseen molecules, provided that they have a local chemical structure similar to the ones in the training set.

A model generating fast predictions for the density matrices provides, first of all, a way to accelerate KS-DFT calculations by generating a better starting guess for self-consistent iterations. Besides the straightforward application to \textit{ab initio} molecular dynamics or geometry optimizations, the possibility of extrapolating predictions to unseen molecules provides the opportunity to accelerate KS-DFT calculations on many different molecules without molecule-specific training. The guesses generated by our unified model allow saving about 20\% of the SCF iterations compared to the default guess in most of the cases. 

Secondly, the predicted density matrix can be used in a quantum chemistry code to directly compute multiple properties. We have tested how well this model predicts energy, Mulliken atomic charges, molecular dipole, and atomic forces. The density matrix predictions give significantly better estimates than the standard guess for all these properties, and especially for the energy, even for unseen molecules.

Our model still presents some limitations. First, although the reduction in SCF iterations is significant, a substantial improvement would be necessary to consistently accelerate KS-DFT calculations by at least a factor of two. Second, the energies and forces predicted by our model still do not reach chemical accuracy, which would be needed for direct applications. Nonetheless, all models show significant room for improvement, both in the model flexibility and in the choice of the training set, making our strategy promising for both applications, especially thanks to the observed systematic improvability and rigorous methodology.

Overall, our results show that learning the density matrix from descriptors encoding the correct symmetry features represents a promising strategy towards more complete and transferable ML models. 
In particular, the density matrix represents the solution of the KS-DFT equations and thus gives direct access to numerous properties with one single ML model. Further, the model turns out to be transferable to unseen molecular structures, which is a central stepping stone towards this development. 

\section*{Acknowledgements}
The authors thank Filippo Lipparini for performing the calculations with the Gaussian development version.
M.N. and B.S. thank the Deutsche Forschungsgemeinschaft (DFG, German Research Foundation) for supporting this work by funding - EXC2075 – 390740016 under Germany’s Excellence Strategy. We acknowledge the support by the Stuttgart Center for Simulation Science (SimTech). 
L.Z. and M.N. and B.S. acknowledge funding by the Deutsche Forschungsgemeinschaft (DFG, German Research Foundation) - Project number 442047500 through the Collaborative Research Center “Sparsity and Singular Structures” (SFB 1481). 
\appendix

\section{Equivariance of the density matrix}\label{app:equiv}
In this appendix, we discuss the equivariance of the density matrix and the rationality of decomposing it in the way that is illustrated in Figure \ref{Fig:DMStructure}. 

Given a configuration $\bR$, and a set of atomic orbitals $\{\phi_{I\alpha}\}_{I,\alpha}$ with which \eqref{eq:kse} is discretized, where $I$ ranges from 1 to $\Nat$ and $\alpha\in\mathcal{I}_{Z_I}$, an index set depending only on $Z_I$. It is straightforward to see that $N_g = \sum_{I=1}^\Nat \#\mathcal{I}_{Z_I}$ with $\# A$ denoting the cardinality of the set $A$. After the discretization, the solution of \eqref{eq:kse} can be used to approximate the KS orbitals, as 
\[
    \varphi_k(\br;\bR) = \sum_{(I,\alpha)} c_{k,(I,\alpha)}[\bR]\phi_{I\alpha}(\br), \quad k = 1,2,\ldots,\Ne.
\]
where the collection of all $\{c_{k,(I,\alpha)}[\bR]\}$ form exactly the eigenvectors $C_\bR$ in \eqref{eq:kse}. Define the density matrix operator by 
\[
    D(\br,\br';\bR) = \sum_{k=1}^\Ne \varphi_k(\br;\bR)^\ast \varphi_k(\br';\bR),
\]
and denote its diagonal by
\[
    D(\br;\bR) = D(\br,\br;\bR).
\]
Then, the elements of the (discretized) density matrix are given by
\begin{eqnarray*}\label{eq:DIJ}
       {\big(D_{\bR}\big)}_{IJ\alpha\beta} &=& \int_{\mathbb{R}^3} \phi_{I\alpha}(\br)^\ast D(\br;\bR) \phi_{J\beta}(\br)d\br \\
       &=&\int_{\mathbb{R}^3} \phi_{\alpha}(\br-\br_I)^\ast D(\br;\bR) \phi_{\beta}(\br-\br_J)d\br  \\
       &=&\int_{\mathbb{R}^3} \phi_{\alpha}(\br-\br_I)^\ast \tilde{D}(\br;\bR_{IJ}) \phi_{\beta}(\br-\br_J) d\br, \quad 1\le I,J\le \Nat, ~\alpha \in \mathcal{I}_{Z_{I}},~\beta \in \mathcal{I}_{Z_{J}}.
\end{eqnarray*}
In the last line, the whole configuration $\bR$ is shifted to be centred at a proper position, which will be elaborated in more details in Appendix \ref{app:ACEBasis} and $\tilde{D}$ is simply the translated density matrix operator. Consequently, the $(I,J)$-th and the $(I',J')$-th blocks of the density matrix $D_{\bR}$ share the same function form as long as $Z_I = Z_{I'}$ and $Z_J = Z_{J'}$, hence gives the unified form \eqref{eq:DIJ}. In addition, we follow an extremely similar discussion as in Ref.~\cite{Zhang2022} to obtain the equivariance of the density matrix. More specifically, if 
\[
    \mathcal{I}_{Z_I} = \{(n, l, m)\}_{\begin{subarray}{l}
    l\in\{0,1,\ldots,l_I\},\\ 
    n\in\{0,1,\ldots,n_{l}\}, \\
    m\in\{-l,-l+1,\ldots,l-1,l\}
    \end{subarray}
    }
\]
where the index $l$ stands for the angular moment, indicating the type of atomic orbital being used (s,p,d orbitals etc.), $n_{l}$ denotes the number of orbitals of the type $l$, and $m$ is the standard angular index corresponding to $l$, then there holds
\begin{equation}\label{eq:equiv}
    \big(D_{Q\bR}\big)_{IJ} = \boldsymbol{\mathcal{D}}_I(Q) \big(D_{\bR}\big)_{IJ} \boldsymbol{\mathcal{D}}_J(Q)^\ast, \quad \forall Q\in\Oiii.
\end{equation}
Here $\boldsymbol{\mathcal{D}}_\bullet(Q) = \text{Diag}(\{\mathcal{D}^{l}(Q)\}_{l\in\mathcal{L}_\bullet})$ with the ordered tuple
\[
    \mathcal{L}_\bullet = [\![\underbrace{0,\ldots,0}_{n_{0}},\underbrace{1,\ldots,1}_{n_{1}},\ldots,\underbrace{l_{\bullet},\ldots,l_{\bullet}}_{n_{l_{\bullet}}}]\!],
\]
and $\bullet$ standing for $I$ or $J$. Moving one step forward, we have (cf. Figure \ref{Fig:DMStructure}(b) and (c))
\begin{equation}\label{eq:sub_equiv}
    \big(D_{Q\bR}\big)_{IJ}^{ll'} = \mathcal{D}^{l}(Q) \big(D_{\bR}\big)_{IJ}^{ll'} \mathcal{D}^{l'}(Q)^\ast, \quad \forall (l,l')\in\mathcal{L}_I\times\mathcal{L}_J,~Q\in\Oiii, 
\end{equation}
which is the smallest unit of describing the isometric symmetry of the density matrix.

\section{The construction of the ACE basis}\label{app:ACEBasis}
Despite the existence of some other possible ways of construction, all the equivariant ACE bases can be obtained through the following procedure, following Ref.~\cite{Zhang2022}: 
\begin{center}
    1-particle basis $\to$ density projection $\to$ $\nu$-body correlation $\to$ symmetrization. 
\end{center}

\noindent The main differences that distinguish the ACE bases are the construction of the 1-particle basis and the symmetrization, while the latter remains the same throughout this paper (cf. \eqref{eq:equiv} or \eqref{eq:sub_equiv}). The focus is thus on the definition of the 1-particle basis. The 1-particle basis is, as its name suggests, a function applied to a single particle $\sigma$, which we will specify in detail for both the \textit{onsite} and \textit{offsite} cases, respectively, for completeness. Suppose the configuration is given by $\bR = \{\sigma_I\}_{I=1,2,\ldots,\Nat}$, with $\sigma_I = (Z_I,\br_I)$.

\vskip 10pt

\noindent {\bf \textit{Onsite} basis:} the onsite environment $\bR_I$ centering at the $I$-th atom is defined as
\[
    \bR_I = \{(Z_K,Z_I,\br_{KI})\}=:\{\sigma_K\}_{K=1,2,\ldots,\Nat, K\ne I},
\]
where $\br_{KI} = \br_K-\br_I$. The second variable in $\sigma_K$ indicates the center of the system. Given a particle $\sigma_K = (Z_K,Z_I,\br)\in\bR_I$, we define the 1-particle basis for the \textit{onsite} $Z_I$ model as
\begin{equation*} \label{eq:one-particle-on}
    \phi^{\rm on}_v(\sigma_K) := \phi^{\rm on}_{Znlm}(\sigma_K) := \delta_{ZZ_K}P_{nl}(r) Y_{lm}(\hat{\br}) \fcut(r)
\end{equation*}
where $\delta_{ZZ_K}$ is the kronecker delta, $\br = r\cdot\hat{\br}$ with $r = |\br|$, and we have identified the composite index $v \equiv (Znlm)$ with $Z$ standing for some atomic number. The radial cutoff function $\fcut(r)$ ensures that only the nearby atoms are taken into account (cf Figure \ref{Fig:loc_env}), which may take different forms. In this work, we choose

\begin{equation} \label{eq:stand_cutoff}
    \fcut(r;r_{\rm cut,Z_I}) = \begin{cases}(r^2/r_{\rm cut,Z_I}^2 - 1)^2,& r\leq r_{\rm cut,Z_I},\\0, &r>r_{\rm cut,Z_I}.\end{cases}
\end{equation}
Here, the subscript $Z_I$ indicates that the cutoff radius can be made element-dependent, and will sometimes be neglected for simplicity. 

\vskip 10pt

\noindent {\bf \textit{Offsite} basis:}
For the \textit{offsite} interactions, we define the offsite local environment as 
\[
    \bR_{IJ} = \{\sigma_{IJ}\}\cup\{\sigma_K\}_{K=1,2,\ldots,\Nat,K\ne I,J},
\]
where $\sigma_{IJ} = \{(Z_I,Z_J), \br_{IJ}\}$, $\sigma_{K} = \{Z_K,(Z_I,Z_J), \br_{IJ,K}\}$ and 
\[
    \br_{IJ,K} = \br_K - \br_{IJ,\theta}, \text{ with } \br_{IJ,\theta} = \br_J+\theta(\br_I-\br_J),~\theta\in[0,1].
\]
The 1-particle basis is then defined as 
\begin{equation*}  \label{eq:one-particle-off}
   \begin{split}
   \phi_{nlm}^\bb(\sigma_{IJ}) &= P_{nl}(r_{IJ}) Y_{lm}(\hat{\br}_{IJ}) \fcutb(r_{IJ}), \\ 
   \phi_{Znlm}^\ee(\sigma_K) &= \delta_{ZZ_K}P_{nl}(r_{IJ,K}) Y_{lm}(\hat{\br}_{IJ,K}) \fcute(\br_{IJ,K};\br_{IJ}).
   \end{split}
\end{equation*}
Here the cutoff function for the bond can be given analogous to those for the \textit{onsite} basis, as
\begin{equation*}
\fcutb(r;r_{\rm bond, Z_IZ_J}) = \fcut(r;r_{\rm bond, Z_IZ_J}),
\end{equation*}
with $\fcut$ being defined in \eqref{eq:stand_cutoff}. On the other hand, the cutoff function $\fcute$ for the environmental atom is defined as 
\begin{eqnarray*}
    \fcute(\br;r_{\rm cut,Z_I},r_{\rm cut,Z_J})
    = \fcut\big(\big|\br + (1-\theta)(\br_J-\br_I)\big|; r_{\rm cut,Z_I}\big) + \fcut\big(\big|\br + \theta(\br_I-\br_J) \big|; r_{\rm cut,Z_J}\big),
\end{eqnarray*}
where $\fcut$ is defined in \eqref{eq:stand_cutoff}, $\br$ denotes the $\br_{IJ,K}$ element in $\sigma_K$ and $r_{\rm cut,Z_I/Z_J}$, the cutoff radii of the two spheres around both the $I$-th and the $J$-th atom. Throughout this work, we choose $\theta=0.5$.

Given the one-particle basis, we can form the density projection and the $\nu$-correlations for the onsite case as
\begin{align*}
    A_v(\bR_I) &:= \sum_{\sigma\in\bR_I} \phi_v(\sigma), \\ 
    \bA_{\bv}(\bR_I) &:= \prod_{t = 1}^\nu A_{v_t}(\bR_I)
        \qquad \text{for } \bv = (v_1, \ldots, v_\nu), ~\nu = 1, 2, \ldots,
\end{align*}
and for the offsite,
\begin{align*}
    A_v(\bR_{IJ}) &:= \sum_{K \neq I, J} \phi_v^\ee(\sigma_K), \\
    \bA_{\bv}(\bR_{IJ}) &:= \phi^\bb_{v^{0}}(\sigma_{IJ})\cdot \prod_{t = 1}^\nu A_{v^t}(\bR_{IJ})
    \qquad \text{for } \bv = (v_0, \ldots, v_\nu), ~\nu = 1, 2, \ldots.
\end{align*}

Finally, we perform the symmetrization over $\Oiii$, by leveraging an averaged integral
\begin{equation*}
    \bB_{\bv,a}(\bR_\bullet) = -\!\!\!\!-\hspace{-4.5mm}\int_{\Oiii} D(Q) (\bA_{\bv}(Q\bR_\bullet) E_a) D(Q)^\ast dQ,
\end{equation*}
where $\{E_a\}_a$ forms a canonical basis of the matrix space where the density matrices $D$, or a sub-block thereof, lies in.

By design, the $\bB_{\bv,a}$ bases have exactly the same equivariance as (the subblocks) of the density matrix $D$. To simplify the notation, we absorbed the index $a$ into $\bv$ in the main text. 

\section{Properties of the retraction}\label{app:proof1}
In this section, we discuss the retraction operator $\P$ and some of its important properties. Denoting 
\[
    S_{N_g} = \{D\in\mathbb{R}^{N_g\times N_g}:D^T=D,\lambda_{D,N}>\lambda_{D,N+1}\},
\]
where $\{\lambda_{D,N}\}_{N=1}^{N_g}$ represents the eigenvalues of $D$, sorted descendingly, then we have the following proposition.

\begin{proposition}
Let $\P:S_{N_g}\to\mathcal{G}_{N_g}^N$ be the retraction operator defined in \eqref{eq:eigen_retract}, then:
\begin{enumerate}
    \item[(1.a)] for $D_{\bR}$ defined in \eqref{eq:equiv}, there holds
        \[
            \P(D_{Q\bR}) = \mathcal{D}(Q)\P(D_{\bR})\mathcal{D}(Q)^\ast, ~\forall Q\in\Oiii;
        \]
    \item[(1.b)] for any $D\in S_{N_g}$, 
        \[
            \P(D) = \argmin_{\tilde{D}\in\mathcal{G}_{N_g}^N} \|\tilde{D} - D\|_F.
        \]
\end{enumerate}
\end{proposition}

\begin{proof}
We first justify that $\P$ is well-defined. Let $D\in S_{N_g}$, then its eigenvalue decomposition can be written as $D = U\Sigma U^T$, where the unitary matrix $U = [u_1, u_2, \ldots, u_{N_g}]$ consists of $N_g$ orthonormal eigenvectors of $D$, $\Sigma$ is a diagonal matrix containing the eigenvalues of $D$, sorted decreasingly. By a straight calculation, we have
\[
    UE_{N_g}^NU^T = \sum_{i=1}^N u_iu_i^T.
\]
Although $U$ may be non-unique, the result in the RHS of the above equality will not be influenced by the order and signs of the first $N$ orthonormal eigenvectors of $D$, since $\lambda_{D,N} > \lambda_{D,N+1}$. This shows the well-definedness of $\P$, \textit{i.e.}, the image of $D$ via $\P$ is unique regardless of how the eigenvalue decomposition is performed. 

Now we can move on to consider (1.a). Assume $D_{\bR} = U_{D_{\bR}} \Sigma_{D_{\bR}} U_{D_{\bR}}^T$, then $\P(D_{\bR}) = U_{D_{\bR}} E_{N_g}^N U_{D_{\bR}}^T$. By \eqref{eq:equiv}, we have that for all $Q\in\Oiii$
\begin{eqnarray*}
    D_{Q\bR} &=& \mathcal{D}(Q) D_{\bR} \mathcal{D}(Q)^\ast, \notag \\
    &=& \mathcal{D}(Q) U_{D_{\bR}} \Sigma_{D_{\bR}} U_{D_{\bR}}^T \mathcal{D}(Q)^\ast,
\end{eqnarray*}
which means that $D_{Q\bR}$ has an eigenvalue decomposition as above. As a result, 
\begin{eqnarray*}
    \P(D_{Q\bR}) &=& \mathcal{D}(Q) U_{D_{\bR}} E_{N_g}^N U_{D_{\bR}}^T \mathcal{D}(Q)^\ast, \notag \\
    &=& \mathcal{D}(Q) \P(D_{\bR})\mathcal{D}(Q)^\ast, \quad \forall Q\in\Oiii.
\end{eqnarray*}
This proves (1.a).

As of (1.b), we again suppose $D = U\Sigma U^T \in S_{N_g}$, with $U$ defined as above. We claim that 
\begin{equation*}\label{eq:best_estimation}
    \P(D) = UE_{N_g}^NU^T = \argmin_{G\in\mathcal{G}_{N_g}^N} \|G - D\|_F = \argmin_{\{PE_{N_g}^NP^T:~P\in\mathcal{O}(N_g)\}} \|PE_{N_g}^NP^T - D\|_F.
\end{equation*}
Note that the last equality above is based on 
\[
    \mathcal{G}_{N_g}^N = \{PE_{N_g}^NP^T:~P\in\mathcal{O}(N_g)\}.
\]
We now estimate
\begin{eqnarray}\label{eq:norm_expansion}
    \|PE_{N_g}^NP^T - D\|_F^2 = \|PE_{N_g}^NP^T - U\Sigma U^T\|_F^2
    = \text{tr}(PE_{N_g}^NP^T + U\Sigma^2U^T - 2PE_{N_g}^NP^TU\Sigma U^T).
\end{eqnarray}
To minimize \eqref{eq:norm_expansion}, we just need to maximize $\text{tr}(PE_{N_g}^NP^TU\Sigma U^T)$, since the trace of the first two terms in the right hand side is a constant, thus
\begin{eqnarray*}
    \max_{P\in\mathcal{O}(N_g)}\text{tr}(PE_{N_g}^NP^TU\Sigma U^T) &=& \max_{P\in\mathcal{O}(N_g)}\text{tr}(U^TPE_{N_g}^NP^TU\Sigma) \\
    &=& \max_{P\in\mathcal{O}(N_g)}\text{tr}(PE_{N_g}^NP^T\Sigma) \\
    &=& \max_{G\in\mathcal{G}_{N_g}^N}\text{tr}(G\Sigma)
    = \max_{G\in\mathcal{G}_{N_g}^N}\sum_{i=1}^{N_g}\sigma_iG_{ii}.
\end{eqnarray*}
For any $G_0\in\mathcal{G}_{N_g}^N$, there exists $P_0\in\mathcal{O}(N_g)$ such that $G_0 = P_0E_{N_g}^NP_0^T$. Consequently, 
\[
    0\le G_{0,ii} = \sum_{t=1}^N P_{0,it}^2 \le \sum_{t=1}^{N_g} P_{0,it}^2 = 1.
\]
In addition, we have $\text{tr}(G_0) = N$. Hence $\frac1N\sum_{i=1}^{N_g}\sigma_iG_{0,ii} = \sum_{i=1}^{N_g}\sigma_i \frac{G_{0,ii}}{N}$ becomes a convex combinition of $\{\sigma_i\}_{i=1}^{N_g}$, which achieves its maximum at $\sum_{i=1}^N \sigma_i$ when $G_0$ is chosen to be $E_{N_g}^N$. This completes the proof. 
\end{proof}


\section{Comparison of fitting the density matrix and the KS matrix}\label{app:DMvsHM}
The KS matrix $F_\bR$ has the same structure and symmetry as the density matrix $D_\bR$ and the learning of it has been more broadly studied compared to that of the density matrix. In this appendix, we compare the fitting of the two objects using the same method. To make our comparison meaningful and fair, we first fix an ACE basis ${\bf B}_{\nu,d}$, where $\nu$ and $d$ stand for the correlation order and polynomial degree that define the basis (in addition, the cutoffs are also fixed but is not explicitly written here for the sake of simplicity). That said, the only thing different for the two targeting models (for the KS matrix and the density matrix) is their coefficients. We then pick the same data points, which have the form $\{(\bR^{(k)}, F_{\bR^{(k)}}, D_{\bR^{(k)}})\}_{k=1}^{N_{\mathrm{data}}}$. By solving the least squares problems \eqref{eq:LS} with the Kohn-Sham matrices and density matrices, respectively, we obtain  $\mathbf{c}_{\nu,d,H}$ and $\mathbf{c}_{\nu,d,D}$ for the two objects. Then we have two routines to get the predicted density matrix of a given configuration $\bR$, which lies in the desired manifold \eqref{eq:Grassmann}.

First, with ${\bf B}_{\nu,d}$ and $\mathbf{c}_{\nu,d,D}$, we obtain directly a feasible approximation of the density matrix
\[
        \tilde{D}_{\bR} = \mathcal{P}(\mathbf{c}_{\nu,d,D} \cdot {\bf B}_{\nu,d}(\bR)),
\]
where $\mathcal{P}$ is the retraction operator defined in \eqref{eq:eigen_retract}. Alternatively, we can construct 
\[
        \tilde{F}_{\bR} = \mathbf{c}_{\nu,d,F} \cdot {\bf B}_{\nu,d}(\bR),
\]
following by solving 
\[
        \tilde{F}_{\bR}\tilde{C}_{\bR} = \tilde{C}_{\bR}E_{\bR}
\]
for its N eigenvectors $\tilde{C}_{\bR}$, and finally obtain
\[
        \bar{D}_{\bR} = \tilde{C}_{\bR}\tilde{C}_{\bR}^T.
\]

\begin{table}[]
\resizebox{\textwidth}{!}{%
\begin{tabular}{c|cc|cc}
\hline
\multirow{2}{*}{\textbf{Molecule}} &
  \multicolumn{2}{c|}{\textbf{Density Matrix}} &
  \multicolumn{2}{c}{\textbf{KS Matrix}} \\ \cline{2-5} 
 &
  \multicolumn{1}{c}{\textbf{Specific Model}} &
  \textbf{Unified Model} &
  \multicolumn{1}{c}{\textbf{Specific Model}} &
  \textbf{Unified Model} \\ \hline
Acetaldehyde   & 4.416$\cdot$10$^{-5}$ & 3.278$\cdot$10$^{-4}$ & 2.369$\cdot$10$^{-5}$ & 2.137$\cdot$10$^{-4}$ \\
Acrolein       & 2.514$\cdot$10$^{-4}$ & 5.028$\cdot$10$^{-4}$ & 1.688$\cdot$10$^{-4}$ & 4.052$\cdot$10$^{-4}$ \\ \hline
Aniline        & 4.300$\cdot$10$^{-4}$ & 4.868$\cdot$10$^{-4}$ & 5.400$\cdot$10$^{-4}$ & 1.316$\cdot$10$^{-3}$ \\
o-Toluidine    & 5.430$\cdot$10$^{-4}$ & 5.962$\cdot$10$^{-4}$ & 4.760$\cdot$10$^{-4}$ & 3.806$\cdot$10$^{-3}$ \\
m-Toluidine    & 5.384$\cdot$10$^{-4}$ & 5.824$\cdot$10$^{-4}$ & 5.401$\cdot$10$^{-4}$ & 4.115$\cdot$10$^{-3}$ \\
Benzene*       & -        & 4.058$\cdot$10$^{-4}$ & - & 4.333$\cdot$10$^{-4}$ \\
Toluene*       & -        & 6.369$\cdot$10$^{-4}$ & - & 4.893$\cdot$10$^{-4}$ \\
Phenol**       & -        & 4.809$\cdot$10$^{-3}$ & - & 3.204$\cdot$10$^{-2}$ \\
Benzaldehyde** & -        & 4.129$\cdot$10$^{-3}$ & - & 3.101$\cdot$10$^{-2}$ \\
p-Toluidine**  & -        & 2.840$\cdot$10$^{-3}$ & - & 1.922$\cdot$10$^{-2}$ \\ \hline
1-Propanol     & 3.049$\cdot$10$^{-4}$ & 4.427$\cdot$10$^{-4}$ & 4.044$\cdot$10$^{-4}$ & 3.468$\cdot$10$^{-4}$ \\
1-Butanol      & 4.510$\cdot$10$^{-4}$ & 4.921$\cdot$10$^{-4}$ & 2.801$\cdot$10$^{-4}$ & 3.579$\cdot$10$^{-4}$ \\
2-Butanol      & 9.173$\cdot$10$^{-4}$ & 1.494$\cdot$10$^{-3}$ & 5.748$\cdot$10$^{-4}$ & 1.709$\cdot$10$^{-3}$ \\
1-Hexanol      & 1.031$\cdot$10$^{-3}$ & 5.324$\cdot$10$^{-4}$ & 3.264$\cdot$10$^{-4}$ & 4.193$\cdot$10$^{-4}$ \\
Ethanol*       & -        & 9.644$\cdot$10$^{-4}$ & - & 1.133$\cdot$10$^{-3}$ \\
2-Propanol*    & -        & 7.701$\cdot$10$^{-4}$ & - & 2.948$\cdot$10$^{-3}$ \\
2-Hexanol*     & -        & 7.384$\cdot$10$^{-4}$ & - & 1.219$\cdot$10$^{-3}$ \\
1-Heptanol*    & -        & 8.896$\cdot$10$^{-4}$ & - & 2.234$\cdot$10$^{-3}$ \\ \hline
\end{tabular}%
}
\caption{The average test set RMSEs on the density matrices obtained by the (3,8) specific models and unified model trained with the density matrices and the KS matrices ($r_{\text{cut}}$ = 4.0).}
\label{tab:DMvsHM}
\end{table}

In table \ref{tab:DMvsHM}, we compare the element-wise test set RMSE in the predicted density matrices obtained by the two approaches above, including both the cases of specific models and the unified model. To be fair, we use exactly the same data points as those mentioned in Section \ref{sec:result} for training. It can be observed that fitting the KS matrix gives comparable, or even smaller test set RMSEs on the predicted density matrices with respect to the specific models, whereas it performs poorly on the unified model. This implies that the approach of fitting the KS matrix only by minimizing the element-wise error is either less transferable or requires more careful weighting for different systems. In any case, it is less robust than the same approach for the density matrix.

\begin{figure}[h!]
    \centering
    \includegraphics[width=0.9\linewidth]{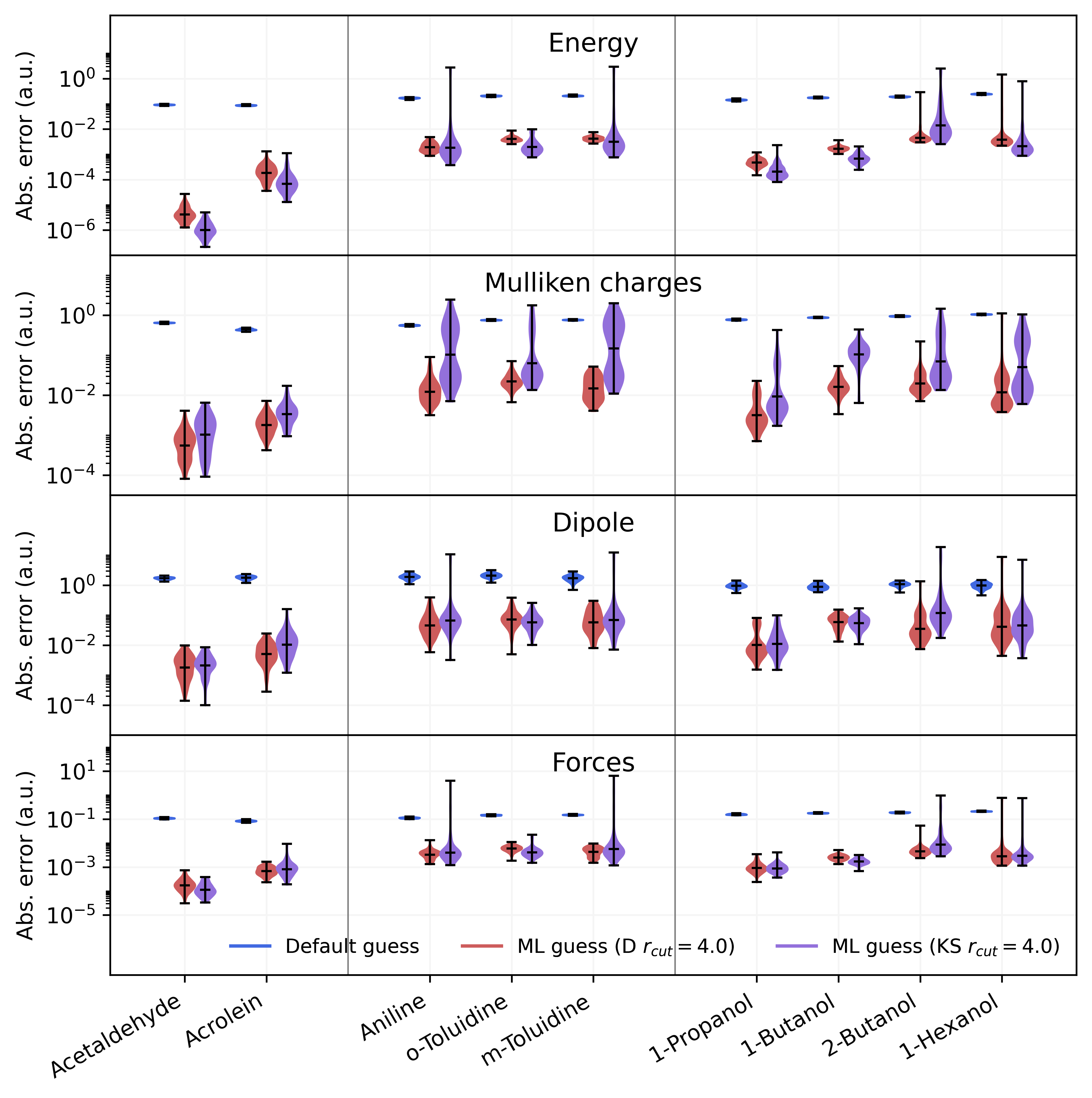}
    \caption{Plot of the error (in logarithmic scale) for energy, Mulliken charges, dipole moment and forces after a single SCF cycle. The blue line represents the default guess provided by Gaussian, and red and violet lines refer to the ML guess obtained with models trained on the density matrices and the KS matrices, respectively ($r_{\text{cut}}$ = 4.0).}
    \label{fig:obs_specific}
\end{figure}

In addition, to see the performance of the specific models more clearly, we compared the properties derived from the specific models tailored to both the density matrices and the KS matrices. The errors on the properties are illustrated in Figure \ref{fig:obs_specific}, from which we can see that although the approach of predicting the KS matrices provides smaller element-wise errors, it cannot guarantee that the derived properties are equally effective. Overall, the proposed approach of fitting the matrix element is more suitable for the density matrix rather than for the KS matrix, at least for the molecular systems that were mentioned in our experiments. 

\printbibliography

\includepdf[pages=-]{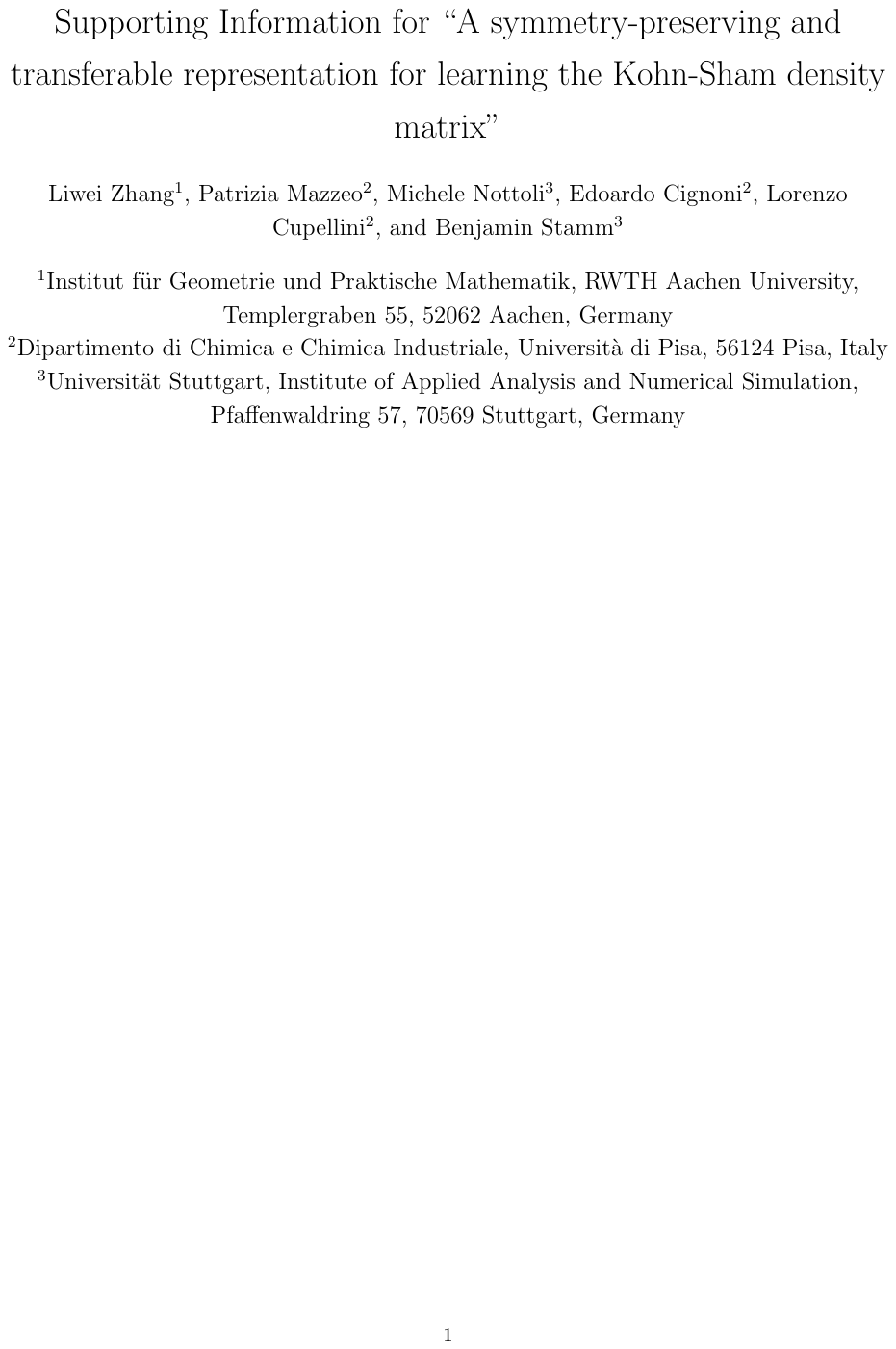}

\end{document}